\documentclass[sigconf]{acmart}
\AtBeginDocument{%
  }

\usepackage{booktabs} 
\usepackage{url}
\usepackage{color}
\usepackage{soul}
\usepackage{booktabs}
\usepackage{graphicx}
\usepackage{multirow}
\usepackage{mathrsfs}
\usepackage{amsbsy}
\usepackage{amsmath}
\usepackage{tcolorbox}
\usepackage{balance}
\usepackage[linesnumbered,ruled]{algorithm2e}

\usepackage{listings}
\usepackage{stackengine}
\usepackage{algpseudocode}
\usepackage{esvect}
\usepackage{xspace}

\newcommand{\bdise}{\textit{\textbf{BDIM-SE}}\xspace}    
\newcommand{\normase}{\textit{\textbf{NorMAS-SE}}\xspace}    

\acmConference[]{}{}{}
\acmBooktitle{}

\begin{document}


\title{Towards autonomous normative multi-agent systems for Human-AI software engineering teams}

\author{Hoa Khanh Dam}
\email{hoa@uow.edu.au}
\affiliation{
  \institution{Decision Systems Lab\\ University of Wollongong}
  \country{Australia}
}

\author{Geeta Mahala}
\email{gm168@uow.edu.au}
\affiliation{
  \institution{Decision Systems Lab\\University of Wollongong}
  \country{Australia}
}

\author{Rashina Hoda}
\email{rashina.hoda@monash.edu}
\affiliation{
  \institution{Monash University}
  \country{Australia}
}

\author{Xi Zheng}
\email{james.zheng@mq.edu.au}
\affiliation{
  \institution{Macquarie University}
  \country{Australia}
}

\author{Cristina Conati}
\email{conati@cs.ubc.ca}
\affiliation{
  \institution{University of British Columbia}
  \country{Canada}
}


\begin{abstract}

This paper envisions a transformative paradigm in software engineering, where Artificial Intelligence, embodied in fully autonomous agents, becomes the primary driver of the core software development activities. We introduce a new class of software engineering agents, empowered by Large Language Models and equipped with \emph{beliefs, desires, intentions, and memory} to enable human-like reasoning. These agents collaborate with humans and other agents to design, implement, test, and deploy software systems with a level of speed, reliability, and adaptability far beyond the current software development processes. Their coordination and collaboration are governed by \emph{norms} expressed as deontic modalities -- commitments, obligations, prohibitions and permissions -- that regulate interactions and ensure regulatory compliance. These innovations establish a scalable, transparent and trustworthy framework for future Human-AI software engineering teams. 

\end{abstract}





\maketitle

\section{Introduction}

The year is 2035. A software engineering (SE) team has just a few human developers, most of whom are in the leadership and creativity roles (e.g., product manager and designer). The rest of the team are \textbf{tens or even hundreds} of virtual \textit{AI-powered, fully autonomous software engineering agents} (\textbf{SE agents}), some specializing in collecting and analysing the customers' needs (business analysts), some in UX/UI designing (designers), and others in programming (coders) and testing (testers). This AI-human team meets regularly to discuss progress and identify blockers, to decide what work to complete, estimate its effort, brainstorm how to accomplish it, assign tasks to team members, review other agents' work, report bugs, and suggest repairs and improvements. SE agents and their human co-workers work seamlessly in concert to continuously build and release software \textit{(says 100 times) faster and more reliably} than today.

This paper aims to realise the above future by introducing a new paradigm for automated software engineering where humans still play a centered role, but the dominant force behind software development is autonomous SE agents. Those SE agents will be fully embedded in the software development process and environment (\emph{situatedness}), operate independently and make their own decisions (\emph{autonomy}), perceive their environment and respond in a timely fashion to changes that occur in it (\emph{reactivity}), have goals that they pursue over time (\emph{pro-activeness}), and interact with humans and other SE agents to accomplish their goals (\emph{social ability}). Central to our approach is the concept of SE teams consisting of multiple specialized autonomous SE agents, each with unique skills and responsibilities. These SE agents work towards achieving a common goal, engaging in collaborative activities like brainstorming, designing, coding, reviewing and testing under human oversight. 


To achieve the above vision, we formulate a roadmap which has the following major research directions: 

\begin{enumerate}
    
    \item \textbf{Research Direction 1:} Design new cognitive architecture specifically for autonomous SE agents that have the key agency characteristics: situatedness, autonomy, reactivity, pro-activeness and social-ability. This
architecture should enable seamless specialization of the
agents into specific SE capabilities such as design, code,
review and test. 

    \item \textbf{Research Direction 2:} Develop a new framework and mechanisms to enable autonomous SE agents to self-regulate their decisions and actions to ensure regulatory compliance with legal and ethical standards.

    \item \textbf{Research Direction 3:} Design and develop a robust teamwork framework for humans and autonomous SE agents in developing software products. This framework will support highly flexible coordination through enabling SE agents to autonomously reason about coordination and communication in Human-AI software engineering teams.

\end{enumerate}

Although recent frameworks such as MetaGPT \cite{hong2024metagpt} and ChatDev \cite{qian-etal-2024-chatdev} demonstrate the feasibility of using Large Language Models (LLMs) \cite{LLMSurvey2023,Gozalo24} to simulate software engineering teams through role-based interactions, they rely on scripted workflows, prompt-engineered roles, and limited adaptability \cite{han2024LLMMASchallenges}. These systems lack genuine agency \cite{He2025,Guo2024}, norm-awareness, and the ability to integrate with human collaborators in real-world settings. By contrast, our research directions represent a paradigm shift towards autonomous SE agents that are cognitively capable, self-regulated and socially aware and integrated, which enables meaningful and accountable collaboration with humans in real-world SE contexts.

\section{\bdise: cognitive architecture for SE agents}
\begin{figure}[ht]
 \centering
    \includegraphics[width=0.45\textwidth]{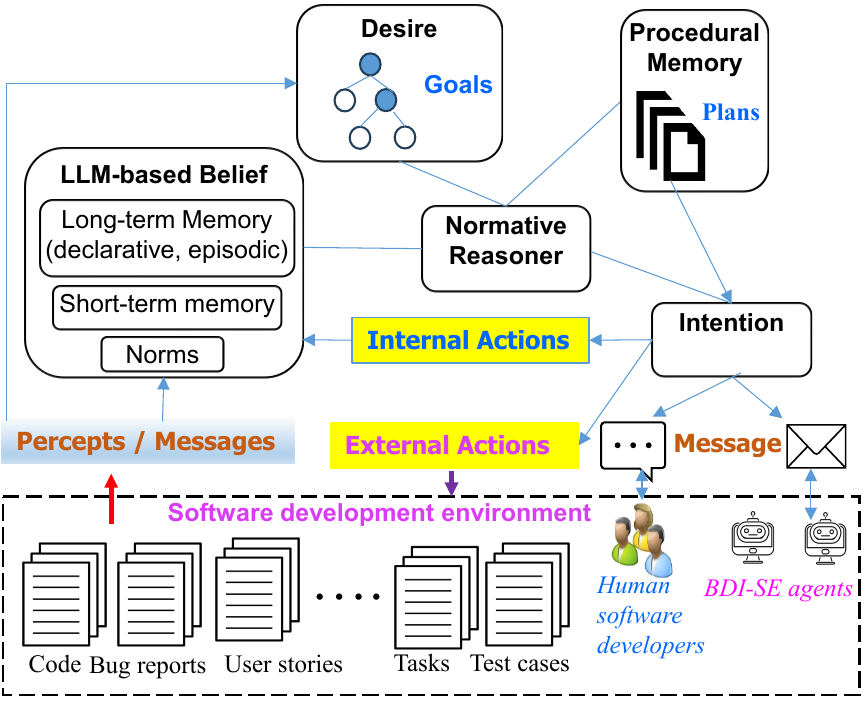}
    \caption{Conceptual architecture of a \bdise agent} 
    \label{fig:arch}
\end{figure}

We propose a new cognitive architecture  (see Figure \ref{fig:arch}) for autonomous software engineering agents with mental attitudes \cite{Silva2020} akin to humans' in terms of \textit{Belief}, \textit{Desire}, \textit{Intention} and \textit{Memory} (\bdise). Each of those components in this cognitive architecture provides a \bdise agent with all the key agency characteristics: situatedness, autonomy, reactivity, pro-activeness and social-ability, which is a significant departure from existing LLM-based agents such as MetaGPT \cite{hong2024metagpt} and ChatDev \cite{qian-etal-2024-chatdev}. Our \bdise agents can autonomously initiate and engage in meaningful interactions with other agents or humans, guided by goals, social norms, and commitments -- capabilities that go well beyond the language-based social skills exhibited by current LLMs \cite{SumersYN024} and automated planning systems \cite{Haslum2019}.  Unlike classical BDI agents \cite{rao1995bdi}, \bdise introduces memory as a core cognitive LLM-powered component, which enables longitudinal reasoning and experience-based learning. These capabilities are crucial for complex software engineering tasks and are often overlooked in traditional cognitive agent models.

Each autonomous \bdise agent will be fully embedded into a software development (\emph{situatedness}), thus through their perception, they have \textbf{beliefs} about all aspects of a software project, its codebase, bug reports, user stories, ongoing tasks, test cases and reports, and customer feedback. For example, our \bdise agent perceives and infers that a bug has been reported, a test case has failed, a new user story has been added to the product backlog, or messages from human developers and other agents. These beliefs are dynamically updated as the agent interacts with different systems in the software development environment such as versioning systems (e.g., GitHub), issue tracking systems (e.g., JIRA) or CI/CD systems (e.g., Jenkins). The perception of events goes beyond tool integration, and requires translating and inferring ambiguous, natural language input into structured symbolic representations.

LLMs are considered stateless since they do not retain information between interactions \cite{Wheeler2025}. Although some recent frameworks simulate memory using large context windows (e.g., Claude \cite{claude3}) or retrieval mechanisms (e.g., Retrieval-Augmented Generation \cite{Lewis2020}), these are not persistent memory in the cognitive sense. In contrast, our \bdise agents store and maintain information internally through their belief structures, which cover both short-term and long-term memory by integrating symbolic representations, external retrieval mechanisms, and LLM-based query capabilities. 
Figure \ref{fig:example} shows an example of a coding agent's belief which stores  the locations of codebase and product backlog of a project: \emph{project\_repo(``https://github.com/codebase.git'')}. The agent's belief also keeps track of the current task, $current\_task(\_)$,  and its status, $task\_status(\_, \_)$. 
The belief also contains explicit rules, e.g.,  $incomplete\_task(T) \mathop{\vcenter{\hbox{$:$}}-}  task\_status(T, adopted) \wedge  task\_status(T, \\changes\_requested)$ indicating that an incomplete task is either just adopted or being changed. Our implementation uses AgentSpeak \cite{AgentSpeak},  an agent-oriented programming language. In AgentSpeak, the belief set consists of a number of \emph{belief literals}, each of which is in the form of a predicate $P$ over the first order terms. 

The belief component of our \bdise agents has not only explicit knowledge representation (in terms of symbolic logic stored in its belief), but also short-term working memory and long-term memory (episodic, semantic and procedural) empowered by a foundational LLM. This LLM-based belief helps fill the knowledge gap that is not captured by explicit logic rules. We implement this as a special literal $query\_LLM$ in our \bdise agent's belief. For example, the context condition of plan P3 in Figure \ref{fig:example} has \textit{query\_LLM(``Is '' + T + `` feasible?'')} which asks LLM if it is feasible to implement a given feature/task $T$. This LLM component has significant zero-shot or few-shot learning capabilities. Neuro-symbolic distillation methods \cite{Delfosse23} can also be leveraged to automatically synthesize belief rules from input-output traces (using human-in-the-loop learning), reducing manual authoring efforts and supporting dynamic agent adaptation based on observed task patterns. This novel approach allows the agent to perform more complex, intelligent tasks, bridging the gap between connectionist approaches (like LLMs) and symbolic AI (which focuses on logic, rules, and reasoning). 

\begin{figure}[ht] \centering
\fbox{\parbox{\linewidth}{
\begin{flushleft}
\texttt{\textbf{\underline{CodingAgent}}}\\
\textbf{Beliefs:}\\
project\_url(``https://github.com/codebase.git'').\\
current\_task(\_).  \textit{// the task selected from the backlog}\\
task\_status(\_, \_). \textit{// tracks progress of the current task}\\
// Explicit symbolic rule\\
incomplete\_task(T) :-   \\
    \hspace{.5cm} task\_status(T, adopted) $\wedge$  task\_status(T, revision). \\
\vspace{0.1cm}

\textbf{Plans:} \\
// Top-level goal is decomposed into several subgoals \hfill (P1) \\
+!complete\_project : true $\leftarrow$ \\    
    \hspace{.5cm} !prepare\_project;  \textit{// subgoal} \\
    \hspace{.5cm} !develop\_feature;  \textit{// subgoal} \\
    \hspace{.5cm} !compile\_and\_test; \textit{// subgoal}  \\
    \hspace{.5cm} !commit\_changes.   \textit{// subgoal} \\

\vspace{0.1cm}

// Prepare the project \hfill (P2)  \\
+!prepare\_project : project\_repo(URL) $\leftarrow$ \\
    \hspace{.5cm} clone\_repo(URL);   \\
    \hspace{.5cm} get\_backlog\_item(T); \\
    \hspace{.5cm} +current\_task(T); \textit{// update the agent's belief} \\
    \hspace{.5cm} +task\_status(T, adopted). \textit{// update the agent's belief} \\

\vspace{0.1cm}

// Implement the current task/feature \hfill (P3)\\
+!develop\_feature : current\_task(T) \& task\_incomplete(T)      \\         \hspace{.5cm}  \& \textbf{query\_LLM(``Is '' + T + `` feasible?'')} $\leftarrow$ \\
    \hspace{.5cm} .concat(``Write code to implement ", T, Prompt); \\
    \hspace{.5cm} \textbf{ask\_LLM(Prompt, GeneratedCode);} \\
    \hspace{.5cm} save\_code\_to\_file(FilePath, GeneratedCode);  \\ 
    \hspace{.5cm} +task\_status(T, implemented). \textit{// update belief} \\

\end{flushleft}
}}
\caption{An example excerpt of \bdise Coding Agent}\label{fig:example}
\end{figure}

The \textbf{desire} component of our \bdise agent consists of software development goals and sub-goals which the agent aims to achieve such as fixing bugs, improving performance, completing user stories or enhancing code quality. The goals are also dynamically updated in responding to changes in the development environment (e.g., a bug found invalid, thus the goal of fixing it is dropped) or coordination with other \bdise agents and/or human developers (e.g., a new goal of completing a user story is adopted after a human developer delegates it to the agent).  The goal-driven approach offers robustness and flexibility, e.g., in case of failures, the agent will try other means to achieve the goal. 

Figure \ref{fig:example} shows an example of the goals and sub-goals of a \bdise coding agent. Here, a top-level goal $complete\_project$ is decomposed into several subgoals: $prepare\_project$, $develop\_feature$, $compile\_and\_test$, and $commit\_changes$. A \bdise agent also has \textbf{intention} for achieving those goals that are realized through their own specific \emph{plans}. Plans consist of external actions or internal actions. External actions can directly change the state of the environment such as running unit tests or refactoring some parts of the codebase or sending messages to other agents. Internal actions interact with internal belief and memories such as reading from long-term memory or updating short-term working memory.  Since the symbolic reasoning allows for abstract problem-solving and planning, our \bdise agents are able to perform multi-step reasoning and goal-directed actions. 

Our \bdise agents have a collection of pre-defined plans stored in the procedural memory module (see Figure \ref{fig:example}). Each plan consists of: (a) an invocation condition which defines a goal the agent aims to achieve, e.g., $+!prepare\_project$ in plan P2; (b) a context condition (usually referring to the agent's beliefs) which defines the situation in which the plan is applicable, e.g., $project\_url(URL)$ in plan P2 or $current\_task(T) \& task\_incomplete(T)$ in P3; and (c)  a plan body containing a sequence of primitive actions and subgoals (which can trigger further plans) that are performed for plan execution to be successful. For example, the coding agent in Figure \ref{fig:example} aims to achieve goal $+!prepare\_project$ by cloning the project repository to its workspace $clone\_repo(URL)$, selecting a task $T$ from the product backlog $get\_backlog\_item(T)$),  and updating its belief about $T$ being its current task $+current\_task(T)$ and the status changed to adopted $+task\_status(T, adopted)$. Our \bdise agent's capability can go beyond the pre-defined actions by using LLMs. For example, plan P3 generates code for a given task by calling an LLM with a relevant prompt $ask\_LLM(Prompt, NewCode)$. 

The decision-making of a \bdise agent consists of the following key steps. Once a goal is adopted, the agent selects from its plan library a set of relevant plans (i.e., matching the invocation condition) for achieving this goal (by looking at the plans' definition). The agent then checks which plan's context condition holds in the current situation to form a set of applicable plans. The agent selects one of the applicable plans and executes it by performing its actions and subgoals.  Execution of a plan, however, can fail in some situation and the agent tries an alternative applicable plan. 
\vspace{-0.4cm}

\section{Human-AI software engineering teams}

In human societies, norms help govern individuals' behavior in group settings (e.g. societies and communities), and regulate the interactions between those individuals \cite{ullmann2015emergence,elster2020social}. Norms dictate which behaviors are encouraged or discouraged, which provides a framework that guides the autonomous decision-making of individuals. Hence, if shared expectations are established, norms can enable autonomous entities, which are either AI agents or humans, to align their actions toward collective goals so that team members behave in ways that support coordination, collaboration, and overall team effectiveness, despite their independence. Therefore, we propose modeling Human-AI software engineering teams as normative multi-agent systems (namely \normase) comprising both human developers and \bdise agents. 

\textbf{Software development norms:} We represent software development norms in terms of deontic modalities \cite{boella2006architecture, garcia2006norm}. \emph{Prohibition norms} prohibit \bdise agents from performing certain actions such as check-in code that violates the human value of privacy, e.g., through a privacy leak. \emph{Obligation norms} describe actions which a \bdise agent is expected to perform such as scanning for security and privacy vulnerabilities before checking-in code. \textit{Permission norms} describe the permissions provided to a \bdise agent such as the team lead agent is allowed to create code branches. There are typically four elements in a norm: \textit{subject}, \textit{object}, \textit{antecedent}, and \textit{consequent}, expressed in JSON format in our implementation. For example, \textit{\{subject: CodingAgent; object: TestingAgent; antecedent: task status(T, implemented); consequent: test(T, ``privacy leaks'')\}} is an obligation norm in which the coding agent expects the testing agent to test if the code implemented for a given task has privacy leaks. Initial norms can be extracted from domain knowledge, ethical AI frameworks, and project governance policies as done in previous work \cite{Dam-ICSE2015,Avery2016}. Human-in-the-loop approaches can also assist in norm authoring, including guided examples and feedback from developers, with potential support from neuro-symbolic distillation methods (e.g., \cite{Delfosse23,howard2024neurocomparatives}). To remain compliant in evolving and multinational contexts of software development, \bdise agents support dynamic norm updates, jurisdiction-aware reasoning, and human-in-the-loop governance, enabling them to adapt to changes in ethical standards, legal requirements, and organizational policies.


\textbf{Self-regulating \bdise agents:} The \textit{\textbf{Normative Reasoner}} component of \bdise agents (see Figure \ref{fig:arch}) internalizes control activities to ensure their behavior are norm-compliant. First, it checks if agent's goals may violate norms (\textit{compliance in desires}) such as a goal entailing a prohibited action or pursuing a goal preventing fulfillment of an obligation. For example, if a coding agent's goal entails storing user inputs, it would detect this violation of a privacy prohibition norm at the goal level, and either reject the goal or revise it. Second,  we extend the BDI intention selection process to select and generate only plans that are compatible with the norms (\textit{compliance in plans}). For example, the coding agent has a goal of implementing user authentication and considers multiple plans, including one that hard-codes API keys. Since this plan violates a norm which prohibits from such practice, \bdise agent filters out that plan and selects a norm-compliant alternative.  Finally, \bdise agent monitors their actions and detect norm violations so that remedies are deployed timely (\textit{compliance in actions}). For example, a \bdise coding agent executes a compliant plan to log system events, however inadvertently logs user IP addresses due to a misconfigured logging library. The agent's runtime monitor detects this norm violation against a privacy prohibition norm and immediately triggers a remedy such as redacting the sensitive data. This normative reasoner also checks the compliance of LLM-generated outputs to prevent LLM hallucination \cite{Huang2025,rawte-etal-2023} and ensure safety, privacy or other norms. Hence, regulatory compliance is achieved \textit{not} by programming correct behavior into the individual \bdise agents, but by introducing norms guiding the behavior, intention and decision making of the agents. 


\textbf{Coordination and communication:} Traditional coordination mechanisms (e.g., task assignments, Kanban boards or CI/CD checkpoints) are designed for human teams and rely heavily on implicit social expectations, rigid workflows and manual interpretation of rules. These assumptions \textit{do not} hold in hybrid teams where AI agents participate as autonomous collaborators. To fill this gap, we conceptualize collaboration in Human-AI SE teams as a social-contract system  \cite{Mahala2023} where humans and \bdise agents coordinate through explicit commitments. Commitments encode obligation, permission and prohibition norms in both machine-interpretable and human-understandable manner to ensure agents' actions are accountable and aligned with team norms. Commitments are formally represented as \textit{C(x, y, r, u)}, where agent \textit{x} commits to agent \textit{y} that if antecedent \textit{r} holds, then consequent \textit{u} will be realized. For example, when the coding agent submits a pull request, the testing agent creates the commitment \textit{C(TestingAgent, CodingAgent, PR\_Submitted, Test\_Results\_Available)}. 

Human developers interact with \bdise agents through natural language. Empowered by LLMs, \bdise agents translate informal instructions into formal commitments and generate human-understandable explanations for their actions. For example, a human team lead tells the coding agent: ``I'll review the pull request in the next 24 hours.'', which is translated into a commitment \textit{C(TeamLead, CodingAgent, PR\_Submitted, Review\_Within\_24h)}. These declarative commitments not only capture the intended social expectation between humans and agents but also enable the automatic generation of an appropriate interaction protocol between the agents \cite{AAAI-20:Clouseau}. Figure \ref{fig:msg} shows an example of an interaction protocol automatically derived from the semantics of the above commitment between the coding and testing agents. 

\begin{figure}[ht] \centering
\fbox{\parbox{\linewidth}{
\begin{flushleft}
\texttt{\textbf{\underline{CodingAgent}}}\\
!+task\_status(T, implemented) $\leftarrow$  \\
        \hspace{.5cm} ?project\_repo(URL); submitPR(URL); \\    
        \hspace{.5cm} .send(TestingAgent, achieve, test(T, ``privacy leaks'')).\\ 
\vspace{0.3cm}
\texttt{\textbf{\underline{TestingAgent}}} \\
// Test the current task/feature \\
!+test(T, Criteria) $\leftarrow$ \\  
        \hspace{.5cm} .concat(``Generate test cases for '', T, Criteria, Prompt); \\
        \hspace{.5cm}  ask\_LLM(Prompt, TestCode); \\
        \hspace{.5cm}  save\_code\_to\_file(Path, TestCode); \\
        \hspace{.5cm}  complile\_and\_test(Path, TestResult); \\
        \hspace{.5cm} +task\_status(T, tested). \\
// Inform Coding Agent the test result\\        
!+task\_status(T, tested) $\leftarrow$  \\    
        \hspace{.5cm} .send(CodingAgent, tell, test\_result(T, Result)).\\                    
\end{flushleft}
}}
\caption{An example of coordination plans automatically generated from a commitment}\label{fig:msg}
\end{figure}

Hence, our approach eliminates the need for manually encoding coordination plans for the teams as currently required in existing multi-agent frameworks such as LangGraph \cite{LangGraph} or CrewAI \cite{CrewAI}. In addition, our framework does not exhaustively predefine or predict every possible failure or deviation in coordination. Remedies can be automatically generated since the semantics of commitments explicitly define violation states which can be linked to secondary commitments that prescribe corrective actions. For example, the violation of the above commitment between the coding and testing agents automatically triggers a remedy in the form of a new commitment (such as reassignment or escalation to a human reviewer) without requiring manual intervention or exhaustive pre-planning. This capability further distinguishes our \normase framework from existing agent orchestration frameworks such as LangGraph and CrewAI (which 
primarily rely on pre-defined workflows or rule-based transitions). Thus, our framework can scale effectively to large and dynamic Human-AI software engineering teams.

\section{Future Plans}

We have implemented the core components of our \bdise agent architecture, integrating a foundational LLM with the belief, desire, intention, and memory modules of individual agents. Our future work will extend the following three complementary directions. 

First, at the \emph{single-agent level}, we will further evaluate the effectiveness of \bdise agents on representative software engineering tasks (e.g., effort estimation \cite{SEEAgent2025}, code generation \cite{jiang2024survey}, testing \cite{WangTSE.2024}, debugging \cite{zhong2024debug}, bug detection \cite{mohajer2023skipanalyzer} and repair generation \cite{Zubair25}). We will compare their performance against state-of-the-art benchmarks using publicly available datasets, and assess their ability to \emph{self-regulate} actions in line with their mental attitudes. 

Second, at the \emph{multi-agent level}, we will formalize and implement normative reasoning within our \normase system, where obligations, prohibitions, and permissions regulate collaboration among multiple agents. We will conduct experiments to measure the added value of \normase compared to single-agent baselines.

Third, at the \emph{Human--AI interaction level}, we will implement commitment-based collaboration mechanisms to structure responsibilities and expectations in mixed teams. We will conduct human user studies with industry practitioners to evaluate how developers perceive \bdise agents as collaborators. Participants will work with the agents on tasks such as effort estimation (e.g., agile planning poker) and code review. To collect feedback from the participants, we will design a questionnaire based on the Technology Acceptance Model framework \cite{davis1989TAM}. We will perform quantitative analysis to assess participants' perceptions of usefulness, ease of use, risks and limitations, and intention towards the use \bdise agents in software development. We will also conduct a qualitative analysis of the observations and open-ended responses. This process involves identifying prominent and recurring themes, patterns, and insights from participants' interactions with agents and their responses.  This mixed-method evaluation will provide a holistic understanding of the usefulness, usability, and acceptance of autonomous SE agents in collaborative software development.  

These investigations will lead to a complete prototype system and a principled empirical foundation for trustworthy, norm-aware, and human-aligned software engineering agents.

\bibliographystyle{ACM-Reference-Format}
\bibliography{references}

\end{document}